\documentclass[final,1p,times]{elsarticle}

\usepackage{amsmath}
\usepackage{amssymb}
\usepackage{amsthm}
\usepackage{rotating}

\newcommand{\vk}{\mathbf{k}}
\newcommand{\vn}{\mathbf{n}}
\newcommand{\vu}{\mathbf{u}}
\newcommand{\vv}{\mathbf{v}}
\newcommand{\vx}{\mathbf{x}}
\newcommand{\vone}{\mathbf{1}}
\newcommand{\vZ}{\mathbf{Z}}

\journal{Mathematical and Computer Modelling}

\begin{document}

\begin{frontmatter}



\title{Huge progeny production during the transient of a quasi-species
model of viral infection, reproduction and mutation}


\author{Jos\'e A.\ Cuesta}
\ead{cuesta@math.uc3m.es}
\ead[url]{http://gisc.uc3m.es/$\sim$cuesta}

\address{Grupo Interdisciplinar de Sistemas Complejos (GISC),
Departamento de Matem\'aticas, Universidad Carlos III de Madrid,
Avenida de la Universidad 30, 28911 Legan\'es, Madrid, Spain}

\begin{abstract}
Eigen's quasi-species model describes viruses as ensembles of different
mutants of a high fitness ``master'' genotype. Mutants are assumed to have
lower fitness than the master type, yet they coexist with it forming the quasi-species.
When the mutation rate is sufficiently high, the master type no longer survives and
gets replaced by a wide range of mutant types, thus destroying the quasi-species. It
is the so-called ``error catastrophe''. But natural selection acts on phenotypes, not
genotypes, and huge amounts of genotypes yield the same phenotype. An important
consequence of this is the appearance of beneficial mutations which increase the
fitness of mutants. A model has been recently proposed to describe quasi-species
in the presence of beneficial mutations. This model lacks the error catastrophe
of Eigen's model and predicts a steady state in which the viral population grows
exponentially. Extinction can only occur if the infectivity of the quasi-species
is so low that this exponential is negative. In this work I investigate the
transient of this model when infection is started from a small amount of low
fitness virions. I prove that, beyond an initial
regime where viral population decreases (and can go extinct), the growth of the
population is super-exponential. Hence this population quickly becomes so huge that
selection due to lack of host cells to be infected begins to act before the steady
state is reached. This result suggests that viral infection may widespread before the
virus has developed its optimal form.
\end{abstract}


\end{frontmatter}


\section{Introduction}
\label{sec:intro}

It seems that an unavoidable consequence of the increase in complexity
of a system is the appearance of parasites. These are entities able to
exploit backdoors, bypasses, holes\dots{} of the system for their own
benefit, sometimes even at a cost for the system. We see a huge variety
of these parasites in biology, ranging from viruses to humans. Society,
in fact, is one of those complex systems amenable to exploitation by
free-riders (the paradigm of the Public Goods game \cite{groves:1977}
is but one prominent acknowledgment of the existence of this social
parasitism). More recently, the widespread use of computers
and the arrival of Internet has made us witness the emergence
and proliferation of computer viruses, trojans, worms, spam, phising, and
all kinds of forms of parasitism, which flood the web using the same
mechanisms aimed at allowing the transmission of information. Apparently,
whenever a complex mechanism emerges, it is soon invaded by its specific
parasites.

Parasites need not be complex: on the contrary, by being very specific
to a particular mechanism, they are able to do their job with very simple
mechanisms. Paradigmatic among parasites for their extreme simplicity
are viruses. Their success is such that they are the most abundant life
forms on Earth~\cite{Koo06}. Their existence is an unavoidable outcome of
the very evolutionary process. In fact, the most common strategy of RNA
viruses is to have
a very high reproductive rate which yields a wide variety of mutants
\cite{Man06}.
This ensures their fast adaptation to almost any change.

One of the most important challenges in current medical research is
how to fight viruses, and one of the most studied strategies is the
design of therapies able to induce viral extinction. Increasing the
mutation rate has been successful, at least in experiments {\it in vitro,}
but there is no consensus as to why the virus loses infectivity at high
mutation rates~\cite{Wil05,Bul05,Tak07,Man10}. The pioneering work
of Eigen~\cite{Eig71} explains viral extinction through a mechanism
known as error threshold. According to it, the progeny loses
its identity if the mutation rate grows above a given value, which
is inversely proportional to the length of the replicating molecule
---hence putting an upper bound to the complexity of viruses.
This classical theory is currently questioned. The current state of
the art of the evolutionary paradigm contradicts some of the basic
assumptions of Eigen's theory, crucial for the existence of the error
threshold. Alternative mechanisms may lead to viral extinction for
reasons other than this hypothetical error threshold (like the presence
of defective forms of the virus~\cite{Gra05,Ira09}, the competition
induced by geometrical constraints~\cite{Pet04,Bar08,Agu08,Cas08,cuesta:2010},
etc.).

Models of viral evolution need to make simplifying assumptions,
and real virus behavior often deviates substantially from their
predictions~\cite{Eig02}. Current quasi-species models assume high
mutation rates that give rise to heterogeneous populations. This is
consistent with experimental observations. However, one common
approximation is to consider that all new mutations have a deleterious effect
on fitness, thus neglecting beneficial and neutral mutations. This
is true if, as the theory assumes, there is a unique master sequence
of high fitness. But we now know that the genotype-phenotype map
is extremely redundant, and that a huge amount of different sequences
---forming so-called neutral networks \cite{gavrilets:2004}--- yield
phenotypes that perform equally well. The increase in the rate of
beneficial and neutral mutations that this effect brings about
invalidates the classical theory of the error threshold~\cite{Man10}
and calls for alternative models of viral evolution and
extinction.

The aim of this paper is to explore one such model, introduced by
Manrubia et al.~\cite{manrubia:2003}, with special focus on its transient
behavior.

\section{Quasi-species equation}

Evolution is a result of the simultaneous action of three processes:
replication, mutation and selection. Any set of agents undergoing these
three processes evolve ---in the direction determined by selection---
regardless of whether they are biological entities, computer programs,
cultural traits, etc.

Replication is the ability of some agents to produce identical copies
of themselves. Replication is normally a stochastic process, characterized
by a probability distribution $p(k)$, $k=0,1,2,\dots$, representing the
probability that after a replication event ---however we define it---
there are $k$ replicas of the parent agent (including itself). 

The replication process is usually imperfect. Most often errors in
making copies yield invalid individuals (unable to produce further copies);
however sometimes these errors produce valid individuals albeit of a
different type (or species). These kind of altered replications are
referred to as mutations. Mutations create new species and maintain
variability within populations. New species may have modified replicative
abilities, and so a probability distribution $p_i(\vk)$ must be introduced
for each species $i$, where $\vk=(k_1,k_2,\dots,k_s)$ is a vector denoting
the number of offspring of any of the $s$ possible resulting species
that an individual of species $i$ gives rise to.

The replication with mutation of an individual of any of the valid species
generates a Markov process in discrete time known as multi-type branching
process~\cite{kimmel:2002}. The variable characterizing this process is
the population of each species at generation $t$,
$\vZ(t)=\big(Z_1(t),Z_2(t),\dots,Z_s(t)\big)$. The mean value of this
variable $\vn(t)=E[\vZ(t)]$ has the simple evolution equation
\begin{equation}
\vn(t+1)=\vn(t)W,
\label{eq:evolution}
\end{equation}
where $W=(w_{ij})$ is the replication-mutation matrix. The number
$r_i=\sum_jw_{ij}$ denotes the average number of offspring that an
individual of species $i$ produces in a replication event, and $q_{ij}=
w_{ij}/r_i$ is the probability that one of this offspring mutates to
species $j$. So introducing stochastic matrix $Q=(q_{ij})$ (mutation
matrix) and the diagonal matrix $R=(r_i\delta_{ij})$ (replication matrix),
we can factorize $W=RQ$, thus separating the effect of replication and
mutation in the evolution of $\vn(t)$.

The asymptotic behavior of this equation is given by $\vn(t)=\lambda^t
\vu$, where $\lambda$ is the largest eigenvalue of $W$ and $\vu$ a
(positive) eigenvector of its corresponding eigenspace. Population
grows exponentially if $\lambda>1$, or vanishes exponentially if
$\lambda<1$.\footnote{If $\lambda=1$ the process is ``critical'', and
it can be proven to go extinct in finite time with probability
one~\cite{kimmel:2002}.}

We have not considered selection yet. Selection is induced by the
environment, usually through a finite availability of resources for
replication. Selection thus acts on the specific replicative ability
of each species ---modifying the values of $r_i$. When scarcity of
resources affects species equally, all values of $r_i$ are affected
equally. In that case, what determines the fate of each species is its
asymptotic fraction within the population. At generation $t$ the fractions
of population of each species is given by the vector $\vx(t)=\vn(t)/
\vn(t)\cdot\vone$, where $\vone=(1,\dots,1)$. Equation~\eqref{eq:evolution}
then becomes
\begin{equation}
\vx(t+1)=\phi(t)^{-1}\vx(t)W, \qquad \phi(t)=\vx(t)W\vone^{\mathsf{T}}
=\sum_ir_ix_i(t),
\label{eq:quasispecies}
\end{equation}
where we have used the factorization $W=RQ$ and the fact that $Q$ is stochastic
(hence $Q\vone^{\mathsf{T}}=\vone^{\mathsf{T}}$). Function $\phi(t)$ represents
the mean replicative ability of the population at generation $t$.
Equation~\eqref{eq:quasispecies} is referred to as the \emph{quasispecies
equation.}

The steady state of equation~\eqref{eq:quasispecies} is obtained by solving
the eigenvalue problem $\vx W=\phi\vx$, under the constraint $\sum_ix_i=1$,
$x_i>0$, $i=1,\dots,s$. If $Q$ is an irreducible matrix $\phi$ and $\vx$
are respectively the largest eigenvalue and its corresponding (unique)
normalized left eigenvector of matrix $W$~\cite{seneta:2006}.

\section{Error catastrophe}

Eigen proposed the quasi-species equation as a model for the evolution
of prebiotic replicators~\cite{Eig71} which, in the absence of correction
mechanisms, had a high mutation rate and accordingly a short length.
However it has become a paradigm of viral evolution even for much
longer sequences (RNA, DNA, proteins\dots)~\cite{Bul05,Dom06}.
To envisage Eigen's idea we can think of a space of $L$ long sequences,
labelled $i=0,1,\dots,s$. Each position of these sequences can be occupied
by any element of a given set of them (DNA or RNA bases, alleles of genes,
aminoacids\dots). Let us assume that this set contains $a$ elements ($a=4$
for bases, $a=20$ for aminoacids\dots).
Mutations are point-like, i.e., substitutions of the
element at a single position by any other in the set. Thus sequences
ACGGCA and AGGGCA are reached from each other by a mutation, whereas
ACGGCA and AGGGCC are two mutations apart. Any offspring of the replicated
sequence will carry a point mutation with probability $0<\mu\ll 1$.
The sequence labeled as $0$ 
(master sequence) is assumed to have a higher replicative ability (henceforth
\emph{fitness)} than any other sequence. For simplicity, all sequences
are assigned fitness $1$ whereas the master sequence has fitness $f>1$.
We shall denote the fraction of population of the master sequence by $x$.

An important assumption in Eigen's model is that backward mutations that
recover the master sequence are neglected. This is a reasonable assumption
considering that sequences in nature tend to be very long. The master
sequence is recovered with probability $(\mu/D)^h$, where $h$ is the Hamming
distance (number of different positions) between the given sequence and
the master sequence, and $D=(a-1)L$ is the number of point mutants of an
$L$ long sequence.

Under the above assumptions the quasi-species equation~\eqref{eq:quasispecies}
reads
\begin{equation}
x[f(1-\mu D)+\epsilon]=x\phi, \qquad \phi=1+(r-1)x,
\end{equation}
where $\epsilon$ contains those backwards mutation that the theory neglects.
This equation predicts
\begin{equation}
x\approx
\begin{cases}
1-\frac{r}{r-1}\mu D & \text{if\ \ $\mu D<1-\frac{1}{r}$,} \\
0 & \text{if\ \ $\mu D>1-\frac{1}{r}$,} 
\end{cases}
\end{equation}
in other words, if the mutation rate is above a threshold (which decreases
as $L^{-1}$), the master sequence accumulates so many mutations that it
gets lost in a cloud of mutants. This transition is known as the \emph{error
catastrophe} and has provided a line of research to find a therapy against
viral infections based on increasing $\mu$ through the addition of
mutagens~\cite{crotty:2001}.

\section{Phenotype vs.~genotype}

But Eigen's model is fundamentally flaw in the assuming the existence of
a single master sequence or genotype. Biology is extremely redundant. DNA
codes for proteins using a (nearly) universal genetic code based on triplets
of bases or codons. Each codon codes for an aminoacid. But the 64 possible
codons only code for 20 aminoacids plus a STOP signal. In redundant aminoacids,
typically the third base is irrelevant or nearly so. This means that many
mutations changing a base pair in the DNA sequence remain silent when
transcribed into proteins. On their side, proteins fold in a three-dimensional
structure which determines their function. And only a few aminoacids at selected
positions are key to this folding. So the replacement of many of them leaves
the protein structure (hence its function) intact. Evolution can only act
on the macroscopic features of living beings (their phenotype), which are
blind to a huge amount of mutations. In other words, the mapping from
genotypes into phenotypes is from very many to one. The existence of a
master sequence is therefore an entelechy. At most we can only speak of
a master phenotype.

The distribution of genotypes corresponding to a given phenotype on 
genotype space is a rather complicated one. Basically they form so-called
neutral networks~\cite{gavrilets:2004}, i.e., connected components of the
mutation graph through which sequences can be changed by successive mutations
without ever changing the phenotype ---hence their fitness. The most
relevant consequence of the existence of neutral networks is that backwards
(or beneficial) mutations are not negligible, because recovering
the master phenotype (not genotype) is no more an improbable event.
Changing Eigen's model to account for beneficial mutations eliminates
the error catastrophe, as we will see in what follows.

\section{A model with beneficial mutations}

A simple model accounting for the existence of neutral networks has been
recently proposed \cite{manrubia:2003}. In this model, viral phenotypes
are characterized by their replicative abilities, $r\in\{0,1,\dots,R\}$.
The only mutations that the model takes into account are those connecting
neighboring classes (i.e., the effect of a mutation is a slight increase
or decrease in the replicative ability).
An offspring undergoes a deleterious mutation from from class $r$ to class
$r-1$ with probability $p$, and a beneficial mutation from class $r$ to
class $r+1$ with probability $q$. In general it is assumed that $0<q\ll
p\ll 1$.
If we denote $n_r(t)$ the mean number of viral particles in class $r$ at
generation $t$, then
\begin{equation}
\begin{split}
&n_r(t+1)=(1-p-q)rn_r(t)+p(r+1)n_{r+1}(t)+q(r-1)n_{r-1}(t), \qquad
r=1,2,\dots R-1, \\
&n_R(t+1)=(1-p)Rn_R(t)+q(R-1)n_{R-1}(t).
\end{split}
\label{eq:discrete}
\end{equation}
Here $R$ stands for the maximum replicative ability of the virus.
There exists also class $r=0$, with no replicative ability,
whose population is maintained because of deleterious mutations from class
$r=1$. Hence $n_0(t)=pn_1(t)$.

Equations \eqref{eq:discrete} have the form of \eqref{eq:evolution} for
$W=RQ$ with 
\begin{equation}
Q=
\begin{pmatrix}
1-p-q &      q &        &        &     \\
  p   &  1-p-q &     q  &        &     \\
      & \rotatebox{15}{$\ddots$} & \rotatebox{15}{$\ddots$} & \rotatebox{15}{$\ddots$} &     \\
      &        &    p   & 1-p-q  &  q  \\
      &        &        &   p    & 1-p 
\end{pmatrix},
\qquad
R=
\begin{pmatrix}
1 &   &        &  \\
  & 2 &        &  \\
  &   & \ddots &  \\
  &   &        & R
\end{pmatrix}.
\end{equation}
Notice however that $Q$ is only sub-stochastic if we do not include class $r=0$.
This fact may cause the total extinction of the virus. Still, the eigenvalue
equation $\phi\vu=\vu W$ determines the asymptotic behavior of the system
$\vn(t)\sim\phi^t\vu$. Both $\phi$ and $\vu$ are unique because $W$ is irreducible.
Vector $\vu$ normalized as $\vu\cdot\vone=1$ describes the asymptotic fractions
of viral particles in each class ---even in the case that the virus eventually goes
extinct.

For $q=0$ it is easy to check that $\lambda_r=r(1-p)$ and $\vv_r=(v_{r1},\dots,
v_{rR})$, with $v_{rk}=\binom{r}{k}(1-p)^kp^{r-k}$, $r,k=1,\dots,R$, are the
eigenvalues and left eigenvectors of matrix $W$, respectively. Since for every
$p$ the largest
eigenvalue is $\phi=R(1-p)$, we find that $p_c=1-R^{-1}$ defines a
transition value such that the virus proliferates for $p<p_c$ but gets extinct
for $p>p_c$. This transition is similar to Eigen's error catastrophe, except
for the fact that the virus becomes extinct in this case because the lowest fitness
class is $r=0$, unable to infect further cells.

\section{Transient and the infinite classes model}

As for the case $q=0$, for $q>0$ we expect that the largest eigenvalue
$\phi=O(R)$, so a model
with an infinite number of classes will never reach the asymptotic state.
However such a model can be useful to study the initial stages of the
transient behavior provided $R\gg 1$ and initially the population has a low
fitness $r_0\ll R$. The reason is that classes above $r_0$ get populated one
by one, so at least for $0\le t\le R-r_0$ there is no difference between the
model with $R<\infty$ and with $R=\infty$. This is the regime I plan to
analyze here.

So consider that the first of equations~\eqref{eq:discrete} holds for all
$r\in\mathbb{N}$ and assume that $n_r(0)=0$ for all $r>r_0$. Then the
generating function
\begin{equation}
G(z,t)\equiv\sum_{r=1}^{\infty}z^rn_r(t)
\end{equation}
will be a polynomial of degree at most $r_0+t$. Multiplying \eqref{eq:discrete}
by $z^r$ and adding up for all $r\ge 1$ we obtain
\begin{equation}
G(z,t+1)=\left[p+(1-p-q)z+qz^2\right]G_z(z,t)-pn_1(t).
\label{eq:discreteG}
\end{equation}
(Subindexes in functions are meant to denote partial derivatives.)

Let us now introduce the generating functions
\begin{equation}
N_r(s)\equiv \sum_{t=0}^{\infty}\frac{s^t}{t!}n_r(t), \qquad
F(z,s)\equiv \sum_{t=0}^{\infty}\frac{s^t}{t!}G(z,t)
=\sum_{r=1}^{\infty}z^rN_r(s).
\label{eq:NFzs}
\end{equation}
In terms of them equation~\eqref{eq:discreteG} becomes
\begin{equation}
F_s(z,s)=p\left(1+\frac{z}{w_-}\right)\left(1+\frac{z}{w_+}\right)F_z(z,s)-pN_1(s).
\label{eq:PDE}
\end{equation}
where
\begin{equation}
w_{\pm}\equiv\frac{1-p-q\pm \Omega}{2q}, \qquad \Omega\equiv\sqrt{1-2(p+q)+(p-q)^2}.
\end{equation}
The condition for $\Omega$ to be real and positive is $\sqrt{p}+\sqrt{q}<1$. This
condition holds whenever $0<q<p<1/4$. As $p=1/4$ is an extremely high mutation rate,
we shall take for granted that $\Omega\in\mathbb{R}^+$.

The first order partial differential equation~\eqref{eq:PDE} needs to be supplemented
with an initial condition for $F(z,s)$. Indeed, if $\{n_r(0)\}_{r\ge 1}$
is the initial condition of the viral populations, then
\begin{equation}
F(z,0)=G(z,0)\equiv g(z)=\sum_{r=1}^{\infty}z^rn_r(0).
\end{equation}

The characteristic curves of equation~\eqref{eq:PDE} are given by
\begin{equation}
\left(1+\frac{z}{w_-}\right)\left(1+\frac{z}{w_+}\right)^{-1}
e^{\Omega s}=\zeta,
\label{eq:changevar}
\end{equation}
with $\zeta$ an arbitrary constant. We can eliminate $z$ from this equation
to get
\begin{equation}
z=z(\zeta,s)=\frac{p(\zeta-E)}{q(w_+E-w_-\zeta)}, \qquad E\equiv e^{\Omega s}.
\end{equation}
In terms of the variables $(\zeta,s)$ and denoting $f(\zeta,s)=
F\big(z(\zeta,s),s\big)$, equation~\eqref{eq:PDE} becomes
\begin{equation}
-p^{-1}f_s(\zeta,s)=N_1(s), \qquad f(\zeta,0)=g\big(z(\zeta,0)\big)
=g\left(\frac{p(\zeta-1)}{q(w_+-w_-\zeta)}\right),
\label{eq:newPDE}
\end{equation}
whose solution is
\begin{equation}
f(\zeta,s)=g\left(\frac{p(\zeta-1)}{q(w_+-w_-\zeta)}\right)-p \int_0^sN_1(u)\,du.
\label{eq:newPDEsol}
\end{equation}
Substituting \eqref{eq:changevar} into \eqref{eq:newPDEsol} yields
\begin{equation}
\frac{p(\zeta-1)}{q(w_+-w_-\zeta)}=\frac{z+w_+\chi(z,s)}{1-\chi(z,s)}
\equiv\psi(z,s), \qquad
\chi(z,s)\equiv\frac{w_-+z}{w_+-w_-}(E-1).
\end{equation}
It only remains to determine $N_1(s)=F_z(0,s)$. This can be achieved by imposing
$F(0,s)=0$ in \eqref{eq:newPDEsol}, which leads to
\[
p\int_0^sN_1(u)\,du=g\big(\psi(0,s)\big).
\]
Thus the final expression of the generating function $F(z,s)$ is
\begin{equation}
F(z,s)=g\big(\psi(z,s)\big)-g\big(\psi(0,s)\big).
\label{eq:finalFzs}
\end{equation}

\section{Asymptotic behavior of the transient}

Setting $z=1$ in \eqref{eq:NFzs} we get $F(1,s)=\sum_{t=0}^{\infty}
n(t)s^t/t!$, the generating function of the total population of the
virus $n(t)=\sum_{r=1}^{\infty}n_r(t)$. From \eqref{eq:finalFzs},
$F(1,s) =g\big(\psi(1,s)\big)-g\big(\psi(0,s)\big)$, where
\begin{align}
&\psi(1,s)=\frac{1+w_+\chi(1,s)}{1-\chi(1,s)}, &
\chi(1,s)=\frac{w_-+1}{w_+-w_-}\left(e^{\Omega s}-1\right), \\
&\psi(0,s)=\frac{w_+\chi(0,s)}{1-\chi(0,s)}, &
\chi(0,s)=\frac{w_-}{w_+-w_-}\left(e^{\Omega s}-1\right).
\end{align}

Let us assume for simplicity that $g(z)=z^r$, i.e., at time $t=0$ only
a single viral particle of class $r$ is present. We can infer the asymptotic
behavior of $n(t)$ from the singularities of $F(1,s)$~\cite{flajolet:2009}.
There are two sets of singularities: $s_0+i2\pi n_0/\Omega$ and
$s_1+i2\pi n_1/\Omega$, with $n_0,n_1\in\mathbb{Z}$,
which are the solutions to $\chi(0,s)=1$ and $\chi(1,s)=1$, respectively.
In each set, the singularity on the real axis is the one with smallest
modulus, so we shall ignore the remaining ones. Denote
$E_0=e^{\Omega s_0}$ and $E_1=e^{\Omega s_1}$; then
\begin{equation}
E_0=1+\frac{w_+-w_-}{w_-}=\frac{w_+}{w_-} \qquad
E_1=1+\frac{w_+-w_-}{1+w_-}=\frac{1+w_+}{1+w_-}.
\label{eq:E0E1}
\end{equation}
But
\[
\frac{E_0}{E_1}=\frac{w_+(1+w_-)}{w_-(1+w_+)}
=\frac{w_++p/q}{w_-+p/q}\ge 1,
\]
because $w_+\ge w_-$ for all $p,q\ge 0$ (the inequality is strict if at least
one of them is nonzero). Then $s_0\ge s_1$, so $s_1$ is the singularity that
is closest to the origin. From~\eqref{eq:E0E1}
\begin{equation}
\Omega s_1=\log\left(\frac{1+w_+}{1+w_-}\right)
=\log\left(\frac{(1+q-p+\Omega)^2}{4q}\right).
\end{equation}
As
$\lim_{s\to s_1}\frac{d}{ds}\big[1-\chi(1,s)\big]=-\Omega 
\frac{w_-+1}{w_+-w_-}E_1=-q (w_++1)\ne 0$, then
$s_1$ is a simple pole of $\psi(1,s)$. As its residue is $-1/q$, then
\begin{equation}
\psi(1,s)\sim\frac{1}{q }\,\frac{1}{s_1-s}
 \text{\ \ \ as\ \ \ } s\to s_1,
\end{equation}
and therefore
\begin{equation}
F(1,s)\sim \frac{1}{(qs_1)^r}\left(1-\frac{s}{s_1}\right)^{-r}
=\frac{1}{(qs_1)^r}\sum_{t=0}^{\infty}\binom{t+r-1}{t}s_1^{-t}s^t, \qquad
s_1\equiv \Omega^{-1}\log\left(\frac{(1+q-p+\Omega)^2}{4q}\right).
\end{equation}
From this we obtain the asymptotic behavior when $t\to\infty$ of the
total population $n(t)$ as
\begin{equation}
n(t)\sim \frac{1}{q^r}\frac{(t+r-1)!}{(r-1)!}s_1^{-t-r}
\sim A_r\left(\frac{t+r-1}{es_1}\right)^{t+r-1/2}, \qquad
A_r=\frac{1}{(r-1)!q^r}\sqrt{\frac{2\pi e}{s_1}}.
\label{eq:n(t)asympt}
\end{equation}

\section{Discussion}

We have analyzed the transient behavior of Manrubia et al.'s
model~\eqref{eq:discrete} for a very large number of classes ($R\gg 1$),
by transforming it into an infinitely many class model. Although an
explicit solution cannot be found, I have obtained the generating
function associated to the vector of class populations. The singularities
of this function provide the time asymptotic behavior of the total
population of the virus, valid as long as the number of generations
is smaller than $R$. Surprisingly we find that viral population
grows \emph{super-exponentially,} unlike in the steady state.

Analyzing eq.~\eqref{eq:n(t)asympt} more closely, we notice that $s_1$
can have very large values and thus induce an initial decay of the
population. However, this decay gets dominated by the factorial $(t+r-1)!$
as soon as $t>t_d\equiv es_1-r+1$. During this decay time $t_d$ (which
is shorter the larger $r$) fluctuations of the branching process can lead
the virus to extinction. Beyond that interval the virus population starts to
recover and grows at a faster than exponential rhythm.

A standard assumption in studies of viral quasi-species evolution is that
their population is in the exponential
asymptotic state. But if $R\gg 1$ the time to reach this state can be very
long (in fact, it requires at least $R-r$ generations to reach the optimal
class, let alone to attain a stationary distribution among classes). Before
that we have the virus population growing faster than exponential and it is
plausible that resources get exhausted during this transient period. This
means that selection starts playing a role when the steady distribution is
not yet established, leading to a behavior different from what is to be
expected in the asymptotic regime. The effects of this phenomenon are
as yet unexplored.

\section*{Acknowledgements}

I thank Susanna Manrubia for long and useful discussions, and for her
critical reading of the draft. This work is part of two research projects:
MOSAICO, from Ministerio de Educaci\'on y Ciencia (Spain)
and MODELICO-CM, from Comunidad Aut\'onoma de Madrid (Spain).





\bibliographystyle{model1-num-names}


\end{document}